\documentclass[structabstract]{aa}  
\pdfoutput=1
\usepackage{graphicx}
\usepackage[utf8]{inputenc}
\usepackage{natbib}
\usepackage{amsmath}
\usepackage{subfig}
\usepackage{psfig}
\usepackage[dvips]{epsfig}
\usepackage{txfonts}
\begin{document}
   \title{Two-dimensional AMR simulations of colliding flows}

   \author{M. Niklaus\inst{1,3} \and W. Schmidt\inst{2,3} \and J. C. Niemeyer\inst{2,3}
          }

   \institute{Deutsches Fernerkundungsdatenzentrum, Deutsches Zentrum f\"{u}r Luft- und Raumfahrt, Oberpfaffenhofen, Germany \\
			\email{markus.niklaus@dlr.de}
			\and
			Institut f\"{u}r Astrophysik, Universit\"{a}t G\"{o}ttingen, Friedrich-Hund-Platz 1, 37077 G\"{o}ttingen
			\and
			Lehrstuhl f\"{u}r Astronomie, Institut f\"{u}r Theoretische Physik und Astrophysik,
  Universit\"{a}t W\"{u}rzburg, Am Hubland, D-97074 W\"{u}rzburg, Germany
			}


  \abstract
   {Colliding flows are a commonly used scenario for the formation of molecular clouds in 
   numerical simulations. Due to the thermal instability of the warm neutral medium, turbulence is produced by cooling.}
   {We carry out a two-dimensional numerical study of colliding flows in order to test whether statistical properties inferred from adaptive mesh refinement (AMR) simulations are robust with respect to the applied refinement criteria.}
   {We compare probability density functions of various quantities as well as the clump statistics and fractal dimension of the density fields in AMR simulations 
   to a static-grid simulation. The static grid with $2048^2$ cells matches the resolution of the most refined subgrids in the AMR simulations.}
   {The density statistics is reproduced fairly well by AMR. Refinement criteria based on the cooling time or the turbulence intensity appear to be superior to the standard technique of refinement by overdensity. Nevertheless, substantial differences in the flow structure become apparent. }
   {In general, it is difficult to separate numerical effects from genuine physical processes in AMR simulations.}
   {}


   \maketitle
%

\section{Introduction}

Computational fluid dynamics in astrophysics relies on numerical methods that are capable
of covering a huge range of scales. Apart from smoothed particle hydrodynamics \citep{Mona92},
adaptive mesh refinement (AMR) has been applied to variety of problems. This method was
developed by \citet{BerOli84} and \citet{BerCol89}. Among the widely used, publicly available AMR codes for astrophysical fluid dynamics are FLASH \citep{FryOls00}, Enzo \citep{Osh04} and Ramses \citep{Teyss02}. Although there are comparative studies of AMR vs.\ SPH \citep[for example, ][]{SheaNaga05,AgerMoore07,CommHenne08}, the degree of reliance of AMR in comparison to non-adaptive methods has met only little attention so far.

Especially for turbulent flows, it is a non-trivial question whether the solutions obtained from AMR simulations agree with the correct solutions of the fluid dynamical equations at a given resolution level.
For this reason, we systematically compare AMR and static-grid simulations for a particular test problem in this article. We chose a scenario that has been investigated in the context of molecular cloud
formation, namely, the frontal collision of opposing flows of warm atomic hydrogen at supersonic speed \citep{HeitSly06,VazGom07,HenneAud07,HenBan08,WalFol00}. Because of the cooling instability at
densities $\sim 1\,\mathrm{cm}^{-3}$ and temperatures of a few thousand Kelvin, the gas
becomes highly turbulent at the collision interface. Since the instabilities develop on length scales much
smaller than the integral scale, this problem is computationally extremely demanding. The
two-dimensional resolution study of \citet{HenneAud07} showed that the properties of the
turbulent multi-phase medium evolving in these simulations is highly resolution-dependent, and numerical convergence is seen only at resolutions well above $1000^{2}$. In three-dimensional simulations, such high resolutions are infeasible if static grids are used. Consequently, \citet{HenBan08} and \citet{BanVaz08} applied refinement by fixed density thresholds and  
refinement by Jeans mass, respectively, in their three-dimensional high-resolution AMR simulations.

In this article, we consider two-dimensional colliding flows without self-gravity and magnetic fields for a systematic
comparison of AMR simulations to a reference simulation on a static grid. We analyze both
statistical properties and the morphology of the gas fragmentation due to the cooling instability.
This work is organized as follows: In Section 2 the numerical methods are described and the setup of the simulations will be presented in detail. In Section 3, we compare the results from the
different simulations. Section 4 concludes this paper with a summary of the main results and general
remarks on AMR.


\section{Numerical methods and simulation setup}

The simulations presented in this article are accomplished using the open source code Enzo \citep{BryNor97,Osh04}. The compressible Euler equations are solved by means of the staggered grid, finite difference method Zeus \citep{StoNor92I,StoNor92II,StoMih92III}. We included the cooling function ${\fam=2 L}$ defined by \citet{AudHen05} in these equations:
\begin{eqnarray}
 \frac{\partial}{\partial t}\rho + \boldsymbol u\cdot\boldsymbol\nabla\rho &=& 0 \label{equ:euler1}\\
 \frac{\partial}{\partial t}\rho + \boldsymbol\nabla(\rho\boldsymbol u \otimes \boldsymbol u + P) &=& 0 \label{equ:euler2}\\
 \frac{\partial}{\partial t} e + \boldsymbol\nabla[\boldsymbol u(\rho e + P)] &=& -{\fam=2 L}(\rho,T)\label{equ:euler3}.
\end{eqnarray}
The primitive variables are the mass density $\rho$, the velocity $\boldsymbol u$ and the specific total energy $e$ of the fluid. The total energy per unit mass is given by
\begin{equation}
 e = \frac{1}{2}u^2+\frac{P}{(\gamma-1)\rho},
\end{equation}
where $\gamma$ is the adiabatic exponent and the pressure $P$ is related to the mass density $\rho$ and the temperature $T$ via the ideal gas law:
\begin{equation}
 P = \frac{\rho k_B T}{\mu m_H}.
\end{equation}
The constants $k_B$, $\mu$ and $m_H$ denote the Boltzmann constant, the mean molecular weight and the mass of the hydrogen atom, respectively. The gas is assumed to be a perfect gas with $\gamma = 5/3$ and $\mu = 1.4 m_H$. 

   \begin{figure}[t]
   \centering
   \resizebox{\hsize}{!}{\includegraphics[width=0.8\linewidth]{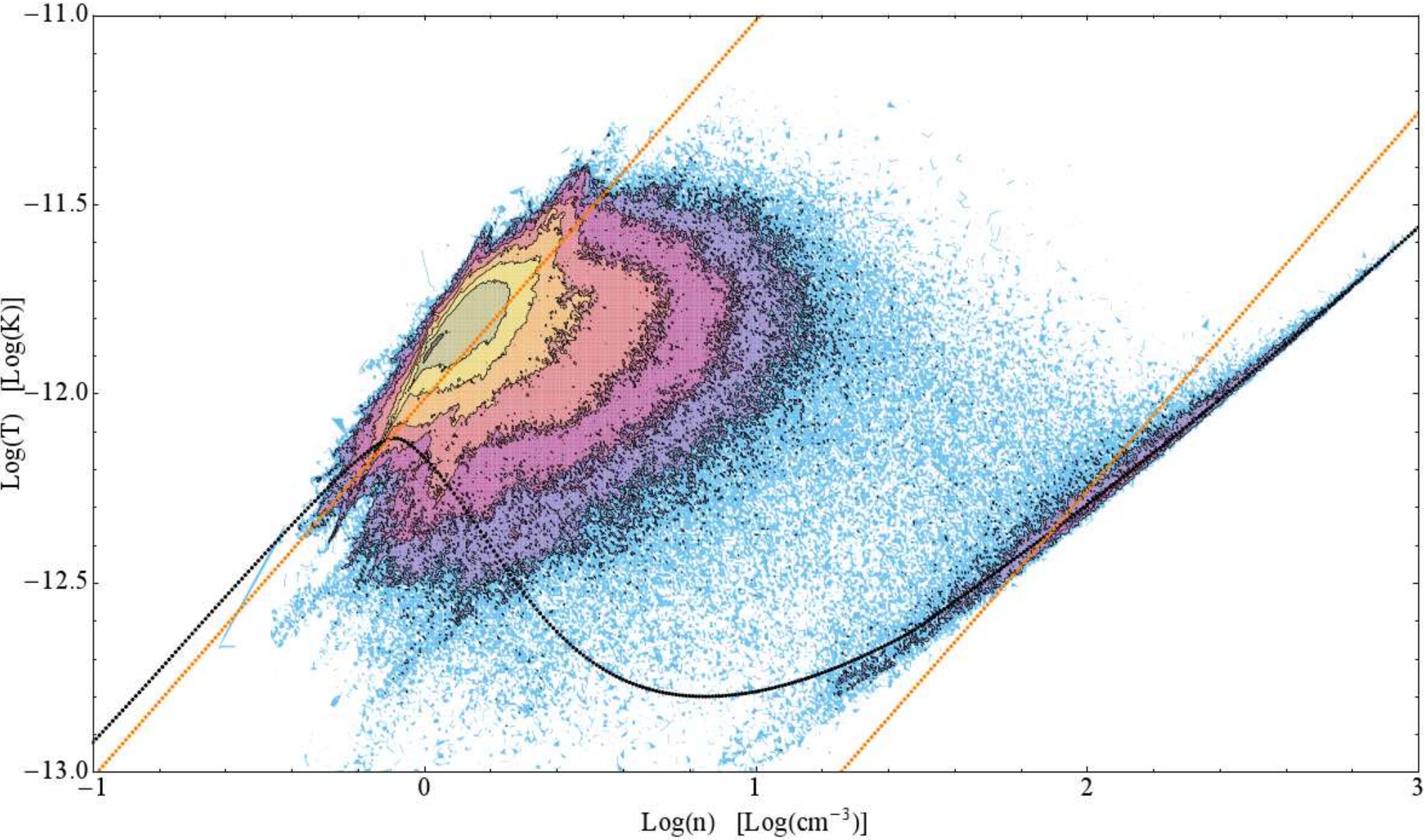}}
      \caption{Phase diagramm of $\log(P)$ vs. $\log(n)$. The solid black curve indicates the equilibrium curve defined by ${\fam=2 L}=0$, and the straight lines (orange in online version) are the isothermals for $7000$ K and $50$ K . The contour shading indicates the probability density.}
         \label{fig:phasediag_static5}
   \end{figure}

The cooling function of \citet{AudHen05} includes the cooling by fine-structure lines of CII and OI, further the cooling by H (Ly$\alpha$-line) and the electron recombination onto positively charged grains. The heating is due to the photo-electric effect on small grains and polycyclic aromatic hydrocarbons (PAH) caused by the far-ultraviolet galactic background radiation. For more information about this cooling function see \citet{Wol95,WolMcK03,Spi78,BakTie94} and \citet{Hab68}. The pressure-equillibrium curve resulting from the cooling function is plotted as black curve in Figure \ref{fig:phasediag_static5}. For the numerical solution of the fluid dynamical equations, we
used the radiative cooling routine implemented into Enzo. For each hydrodynamical time step,
the state variables are iterated over several subcycles, and the resulting total energy increment
for the whole time step is added.

   \begin{figure}[t]
   \centering
   \resizebox{\hsize}{!}{\includegraphics[width=\linewidth]{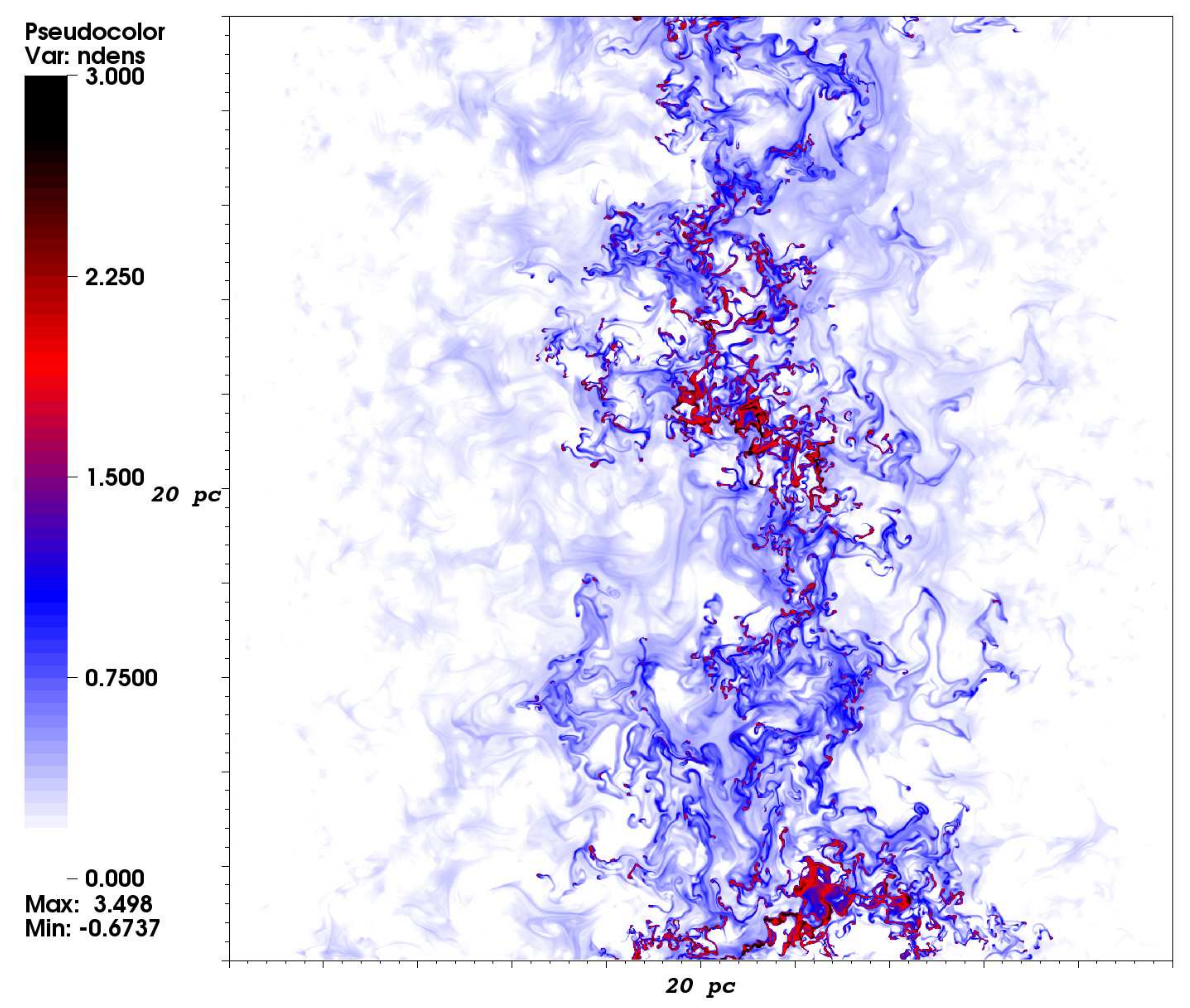}}
      \caption{Contour plot of the mass density after $5$ Myrs of evolution in the static-grid simulation. The density scale is logarithmic.}
         \label{fig:densityplot_static5}
   \end{figure}

For our numerical study, the two-dimensional setup of \citet{AudHen05} and \cite{HenAud07} was adopted with small modifications. The initial gas content corresponds to the warm neutral material (WNM) in the ISM, i.~e., the temperature is $T=7100$ K, the pressure is $P=7\times10^{-13}$ erg cm$^{-3}$ and the number density of neutral hydrogen is $n=0.71$ cm$^{-3}$. From the left and the right boundaries, warm gas with identical thermodynamic properties flows into the computational domain,
where the cosine-shaped inflow velocity profile is modulated with small perturbations, realised by randomly shifted phases.  These phase shifts are kept constant for the different simulations, so that initial conditions are exactly the same for all runs to ensure comparability. Following  \citet{HenBan08}, the top and bottom boundary conditions are periodic. The physical dimensions of the computational domain are $20\times20$ pc. The two inflows of gas collide in the middle of the domain. The supersonic collision causes a steep rise in the gas density that triggers the thermal instability, and gas undergoes a transition into the phase of the cold neutral material (CNM) in the ISM. In this phase, the gas has temperatures in the range $30$--$100$ K and number densities within $20$--$50$ cm$^{-3}$ \citep{Fer01}. The thermal instability produces highly turbulent structures (see Figure \ref{fig:densityplot_static5}) with Mach numbers up to $20$. The challenge for AMR is to track these turbulent structures as accurately as possible. 

   \begin{figure*}[t]
   \centering
   \subfloat[OD]{\label{fig:densityOD}\includegraphics[width=0.3\textwidth]{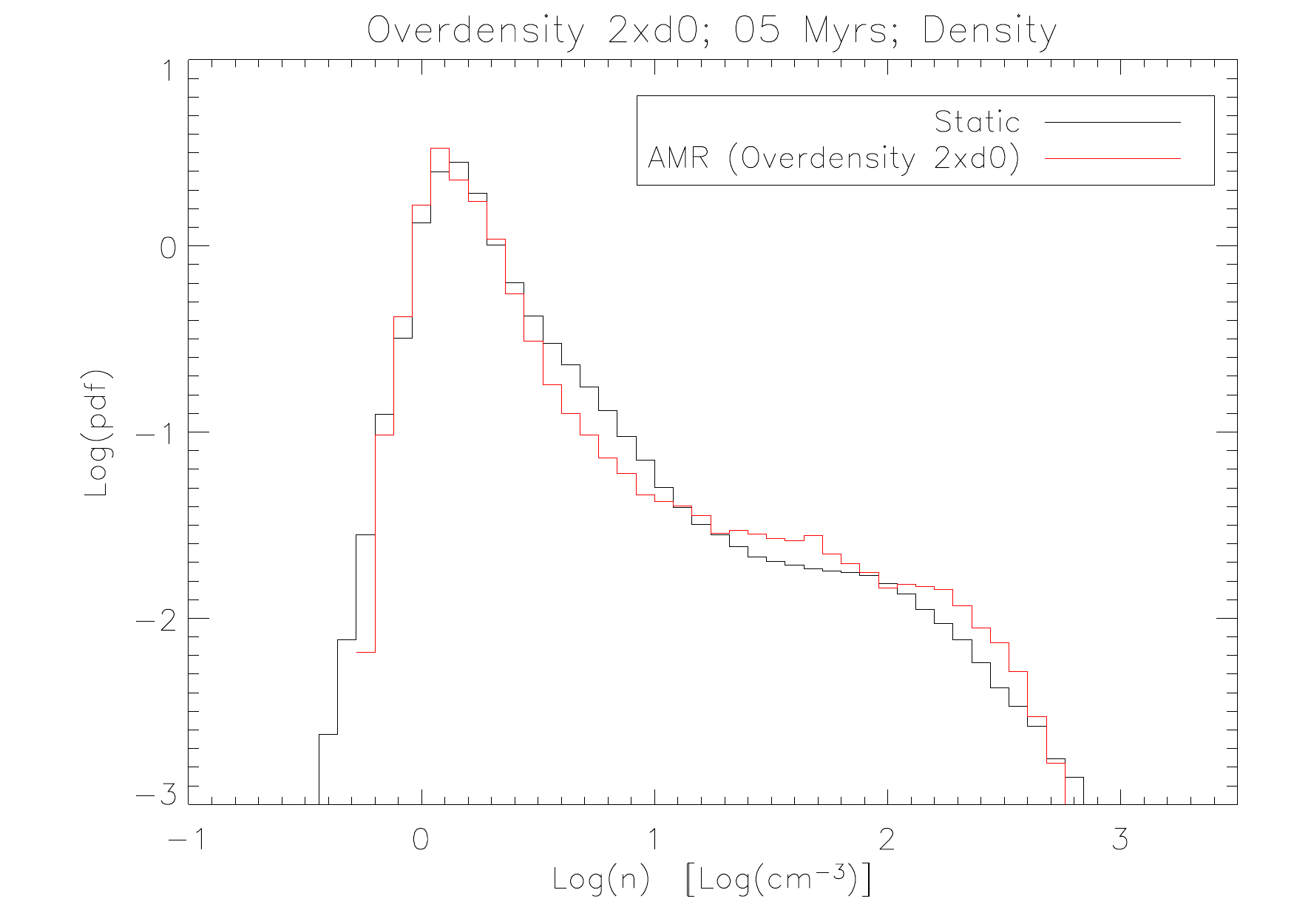}}
   \subfloat[CT]{\label{fig:densityCT}\includegraphics[width=0.3\textwidth]{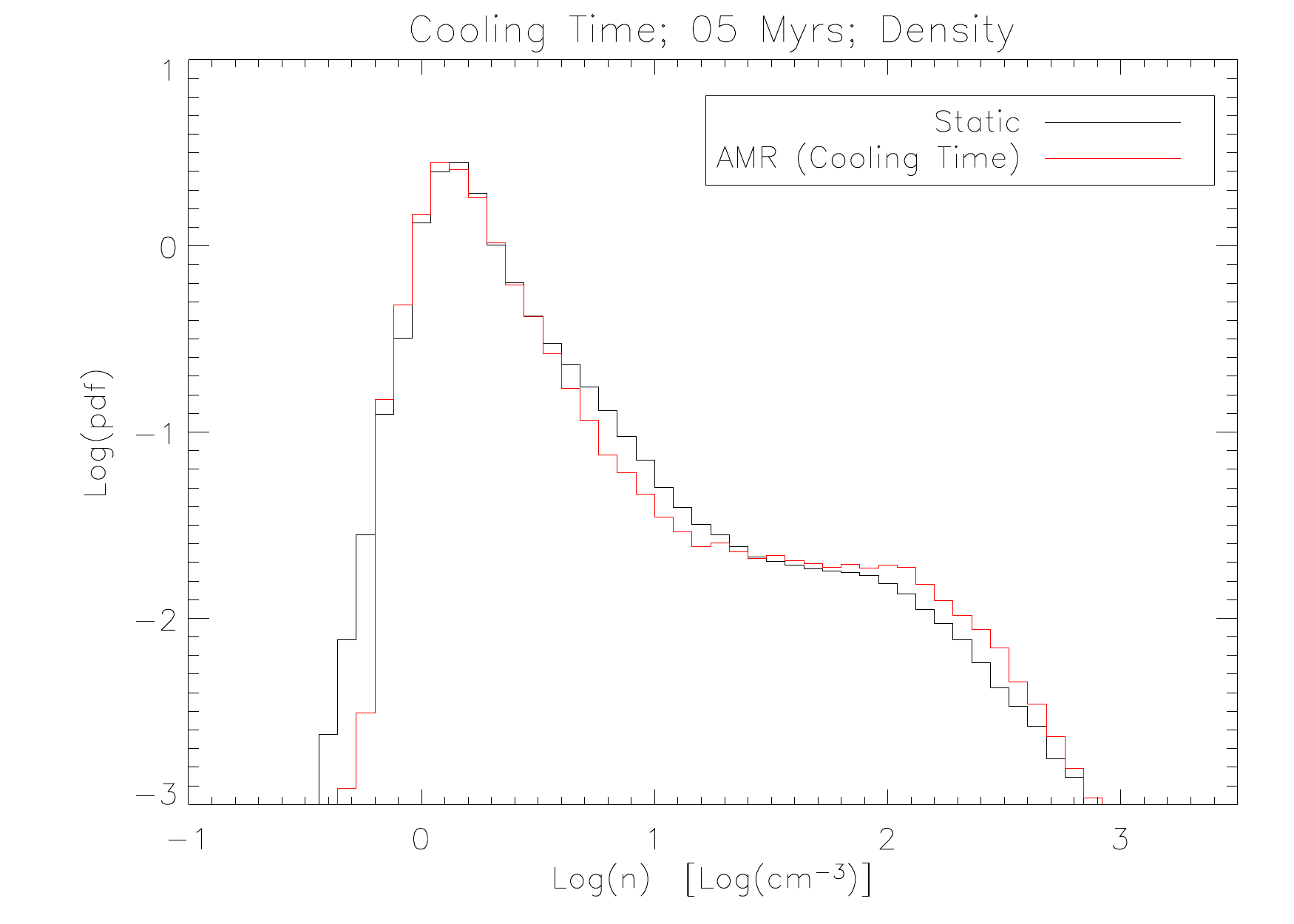}}
   \subfloat[RCEN]{\label{fig:densityRCEN}\includegraphics[width=0.3\textwidth]{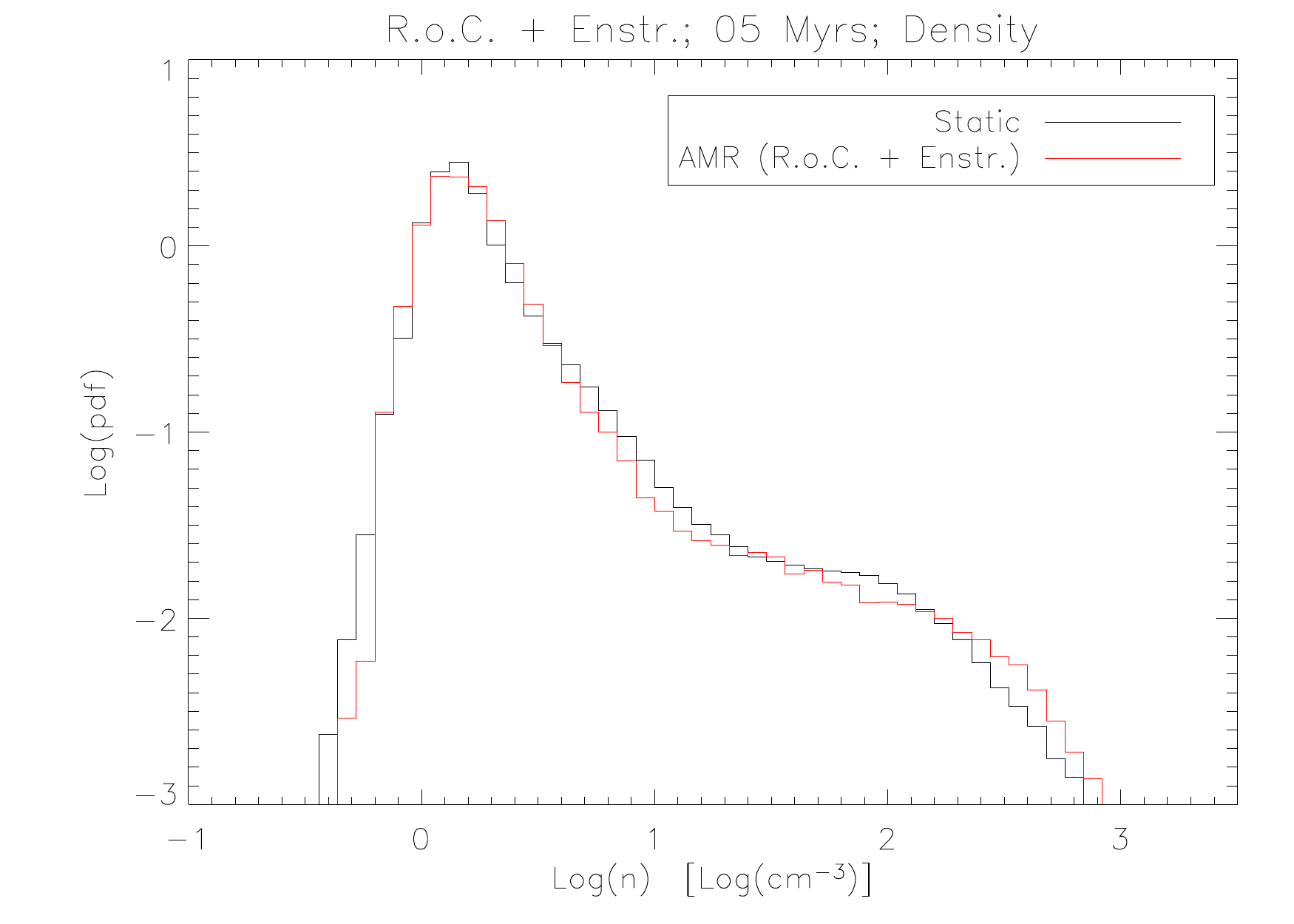}} \\
   \subfloat[OD]{\label{fig:tempOD}\includegraphics[width=0.3\textwidth]{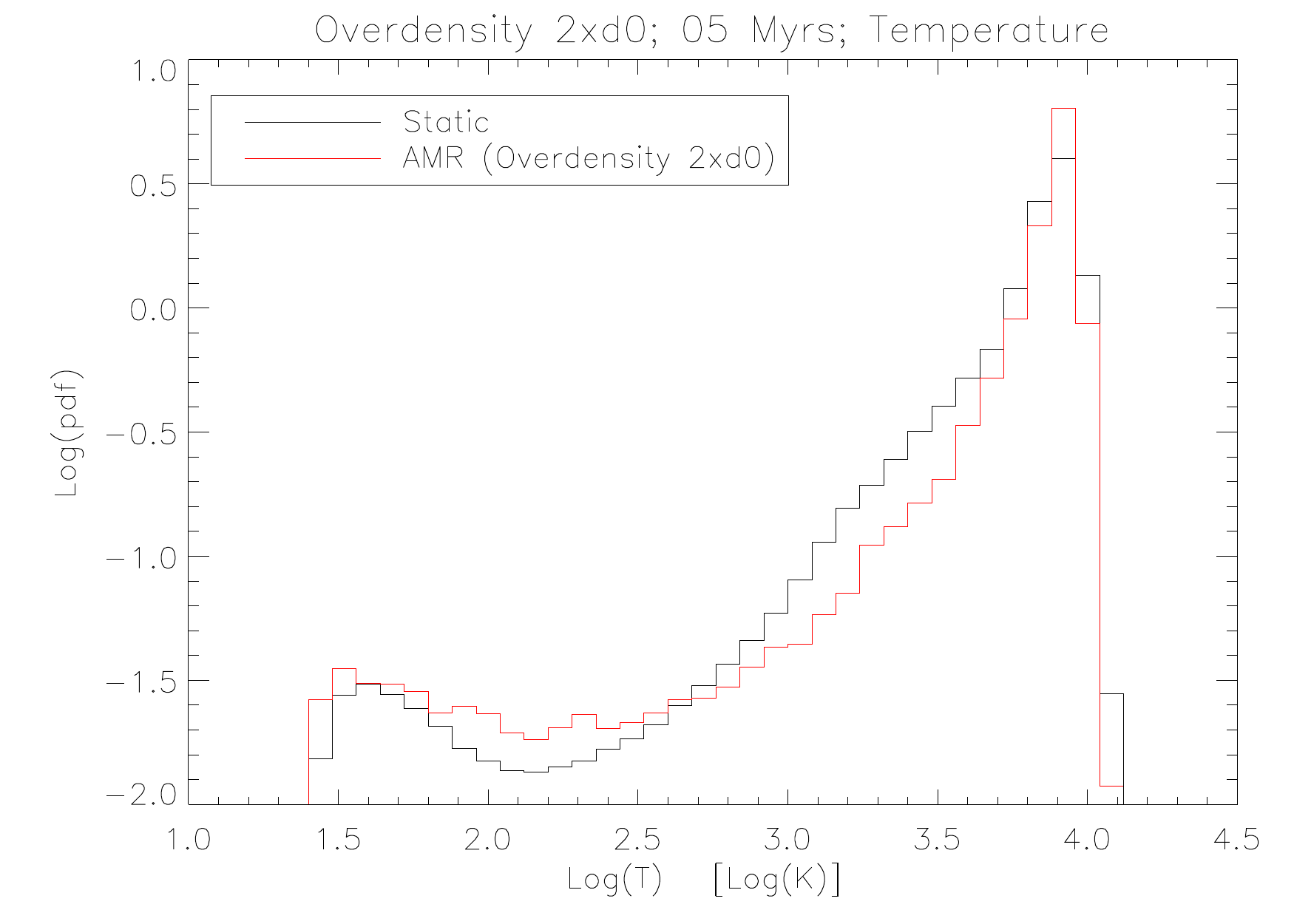}}
   \subfloat[CT]{\label{fig:tempCT}\includegraphics[width=0.3\textwidth]{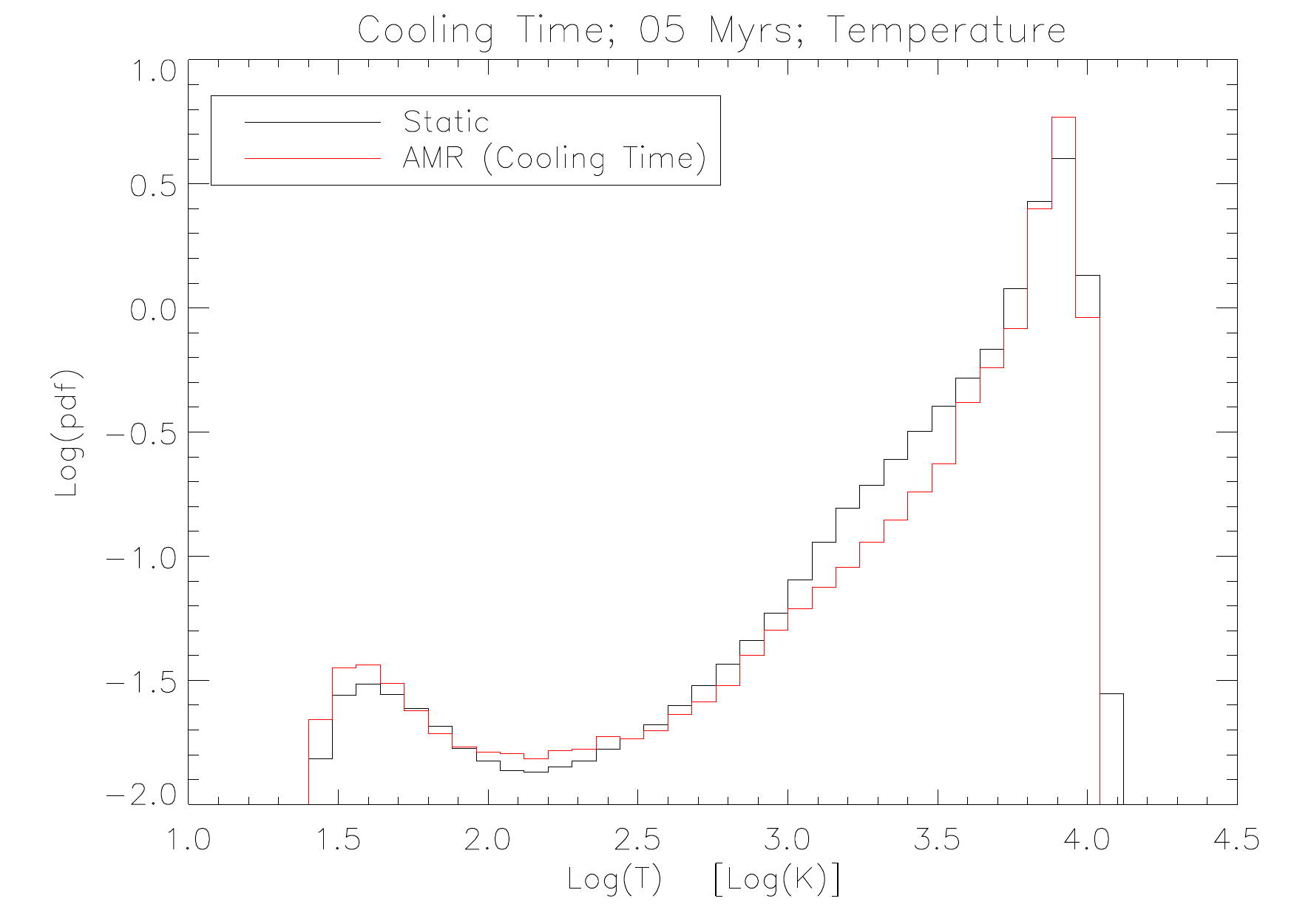}}
   \subfloat[RCEN]{\label{fig:tempRCEN}\includegraphics[width=0.3\textwidth]{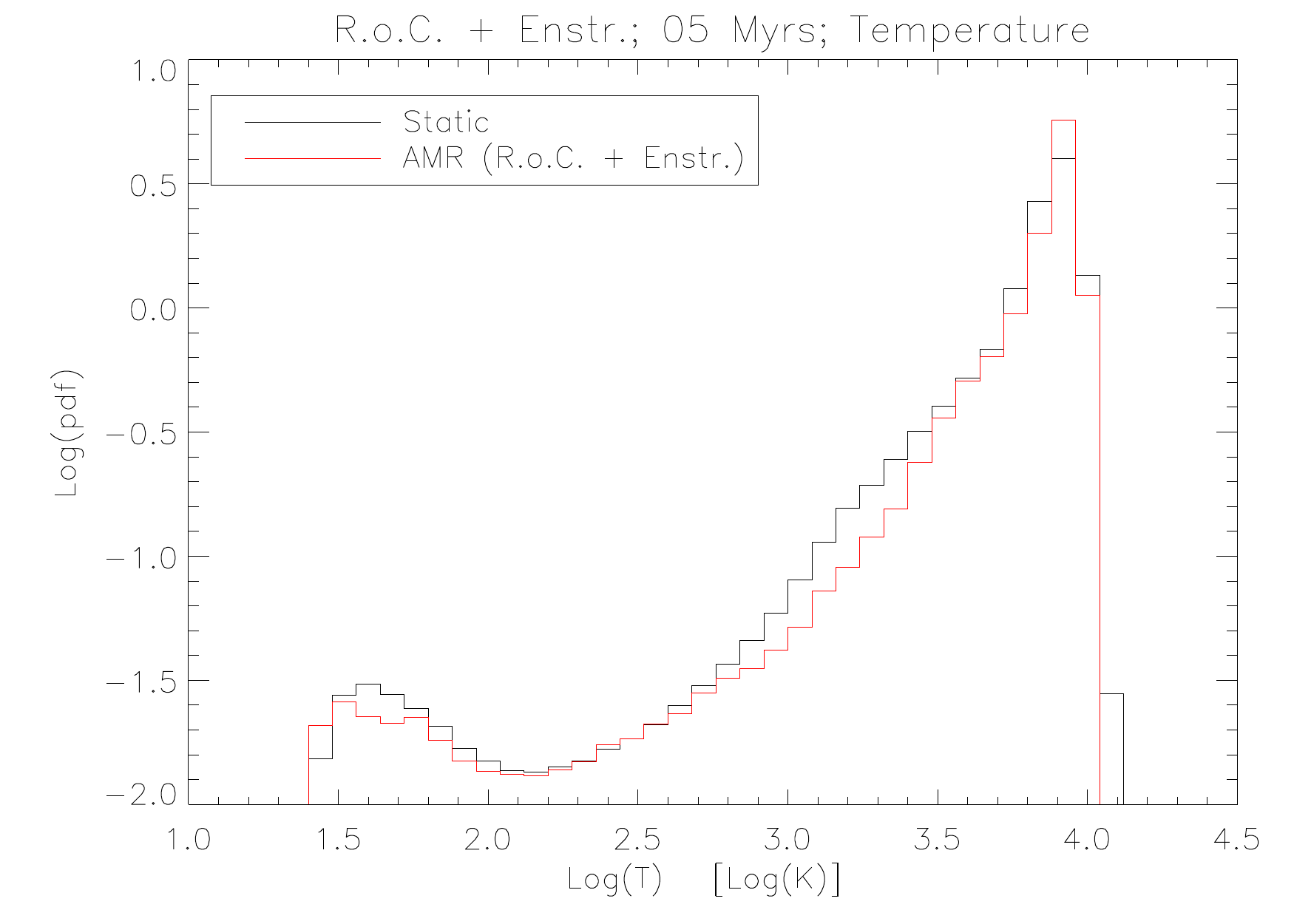}}
      \caption{Pdfs of the number density $n$ (upper panel) and the temperature $T$ (lower panel) in log-log-scale for the different AMR-criteria (black curves) compared to the static grid simulation (red curves in the online version).}
         \label{fig:pdfdenstemp}
   \end{figure*}

A reference simulation was run with a static grid of $2048^2$ cells. Then the same setup was evolved in AMR simulations with a root-grid resolution of $128^2$ cells and $4$ levels of refinement. 
The resolution between adjacent refinement levels increases by a factor of $2$. Hence, the
effective resolution at the highest level of refinement is $2048^2$. In these simulations,
we employed three different types of refinement criteria: 
\begin{enumerate}
\item refinement by overdensity (OD),
\item refinement by cooling time (CT),
\item refinement by rate of compression and enstrophy (RCEN).
\end{enumerate} 
The first two criteria are widely used in astrophysical AMR simulations. For refinement by overdensity,
the mass density must exceed the initial density on the root grid by a certain factor. This overdensity,
in turn, defines the initial density for refinement at the first level of refinement and so on. We
chose three different values for the overdensity factor, namley, twice the initial density (default OD), as well as three times (OD-3) and fourtimes (OD-4) the initial density. 
For criterion CT, on the other hand, refinement is triggered for a grid cell if the cooling time $\tau_{\mathrm{cool}}:=P/(\gamma-1)\rho|{\fam=2 L}|$ becomes less than the sound crossing time over the cell width. Refinement
by the rate of compression and the enstrophy uses yet another technique. It was introduced by
\citet{Sch09} for the simulation of supersonic isothermal turbulence. The control variables for
refinement are the enstrophy and the rate of compression. The enstrophy is given by one-half of the square of the vorticity, while the rate of compression is defined by the substantial time-derivative of the negative divergence of the velocity. The expression used by \citet{Sch09} to evaluate the rate of compression  (see equation (12) in this paper) is easily generalized to the non-isothermal case, where the speed of sound is not a constant. To trigger refinement by RCEN, dynamic thresholds are
calculated from statistical moments of the control variables: A grid cell is flagged for refinement if the local fluctuation of a control variable becomes greater than the maximum of the average and the standard deviation of the variable. On the root grid, averages and variances are computed globally, whereas averaging is constrained to individual grid patches at higher levels of refinement. 

For comparison of the simulation results, we calculated probability density functions (pdf) of several quantities. To analyze the gas fragmentation in each simulation, we adapted the \emph{clumpfind} algorithm implemented by \citet{Pad07} to non-isothermal problems. The algorithm identifies the smallest dense regions that fulfill the Jeans criterion for gravitationally unstable gas. Since the clump
samples found on the two-dimensional grids used in our simulations are insufficient for the calculation
of clump mass spectra, only the total number and the mean size of the clumps are used for quantitative comparisons. In addition, we computed the fractal dimension of gas at densities higher than $n=20\, \mathrm{cm}^{-3}$ (corresponding to the minimum density of gas in the cold phase) by means of the box-counting method \citep{FedKless09}.

\section{Results}

Due to the gradual accumulation of gas in the simulation domain, no strict statistical equilibrium
is approached. For this reason, we evolved the flow until noticeable small-scale structure has
developed and the separation of the gas into two phases has emerged. As shown in Figure~\ref{fig:phasediag_static5}, two distinct phases are found at time $t=5\,\mathrm{Myrs}$. At this time, the central flow region is in a turbulent state (a contour plot of the mass density of the gas is shown in Figure~\ref{fig:densityplot_static5}). Thus, we carry out our analysis for $t=5\,\mathrm{Myrs}$. While the main fraction of the gas is situated in the warm phase with temperatures between $5000$ and $10000$ K and low densities $\sim 1$ cm$^{-3}$, the cold gas with temperatures between $30$ K and $100$ K and densities in the range $30$ -- $350$ cm$^{-3}$ can be found close to the equilibrium curve. 

The pdfs of the mass density and the temperature obtained from different AMR simulations are plotted in Figure \ref{fig:pdfdenstemp}. In principle, all refinement criteria reproduce the distributions found in the static-grid simulation quite well, although there is a trend of slightly more cold gas at the cost of warm gas. The discrepancy is more pronounced for refinement by over-density (OD) compared to the other criteria, and it becomes worse for the higher density thresholds (OD-3 and OD-4; not shown in the Figure). Nevertheless, it appears that the thermodynamic properties of the gas are quite robust in AMR simulations. 

\begin{figure*}[th]
   \centering
   \subfloat[Static]{\label{fig:clumpsstatic}\includegraphics[width=4cm]{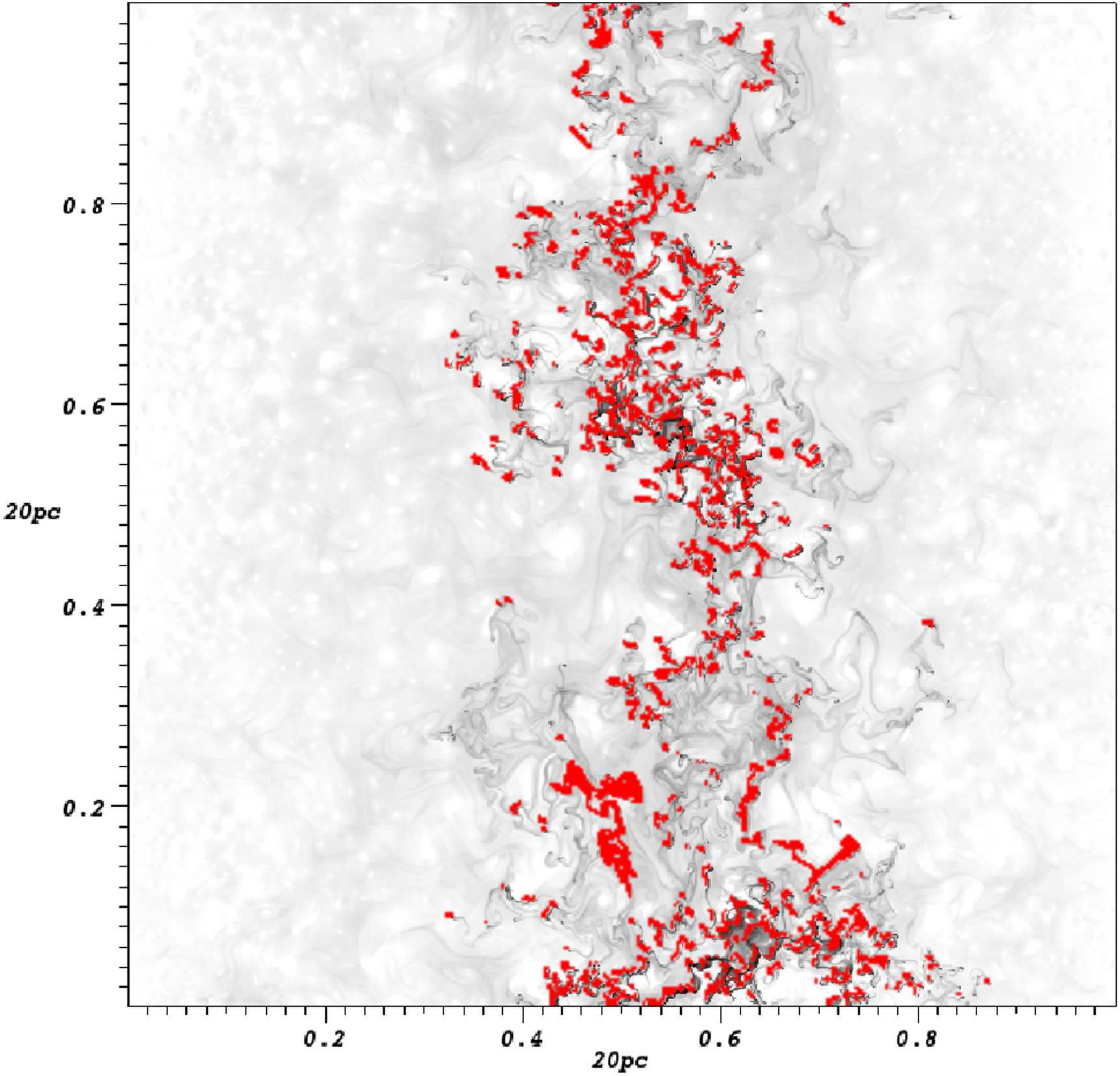}}
   \hspace{5mm}
   \subfloat[OD]{\label{fig:clumpsOD}\includegraphics[width=4cm]{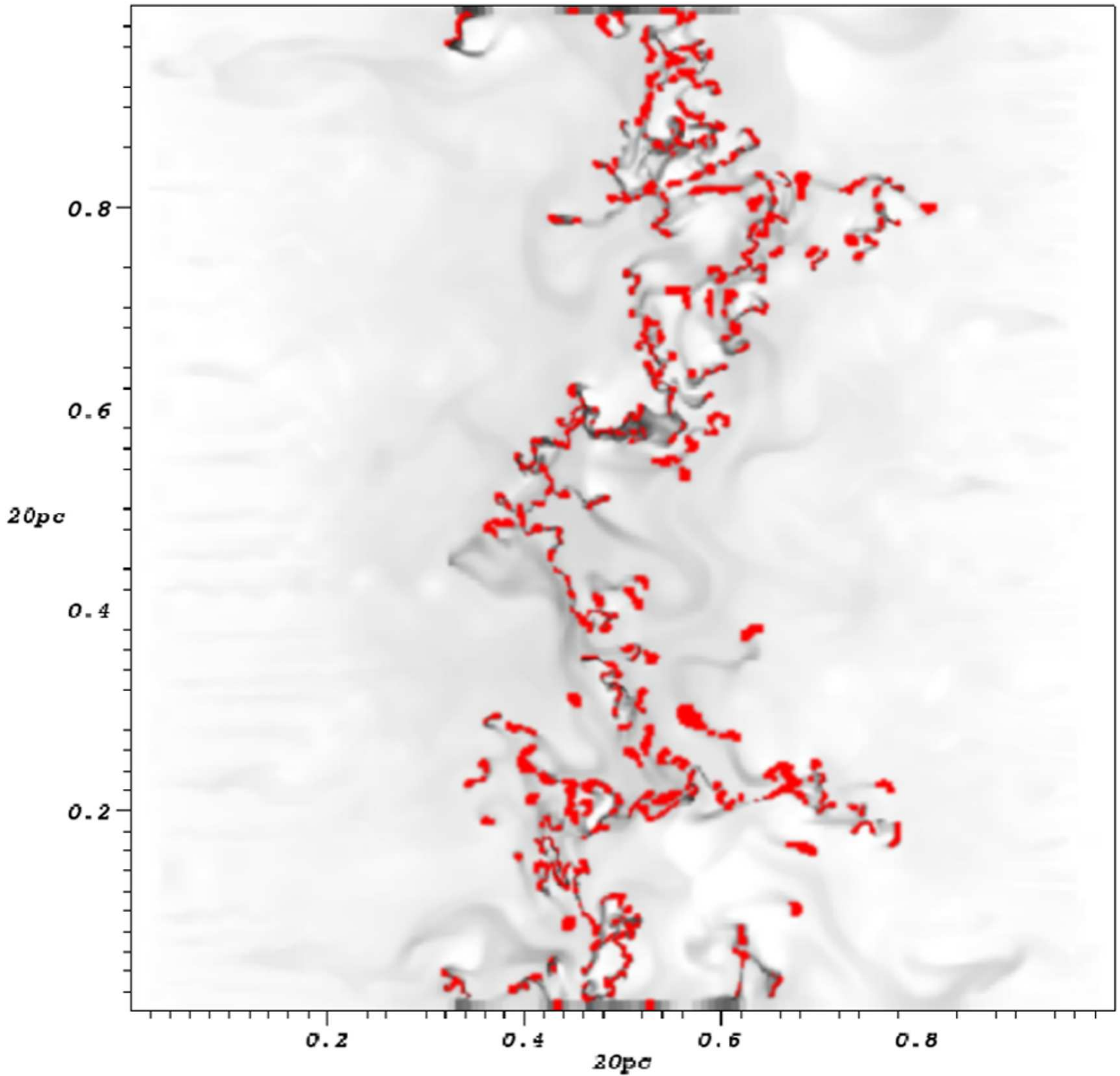}}
   \hspace{5mm}
   \subfloat[CT]{\label{fig:clumpsCT}\includegraphics[width=4cm]{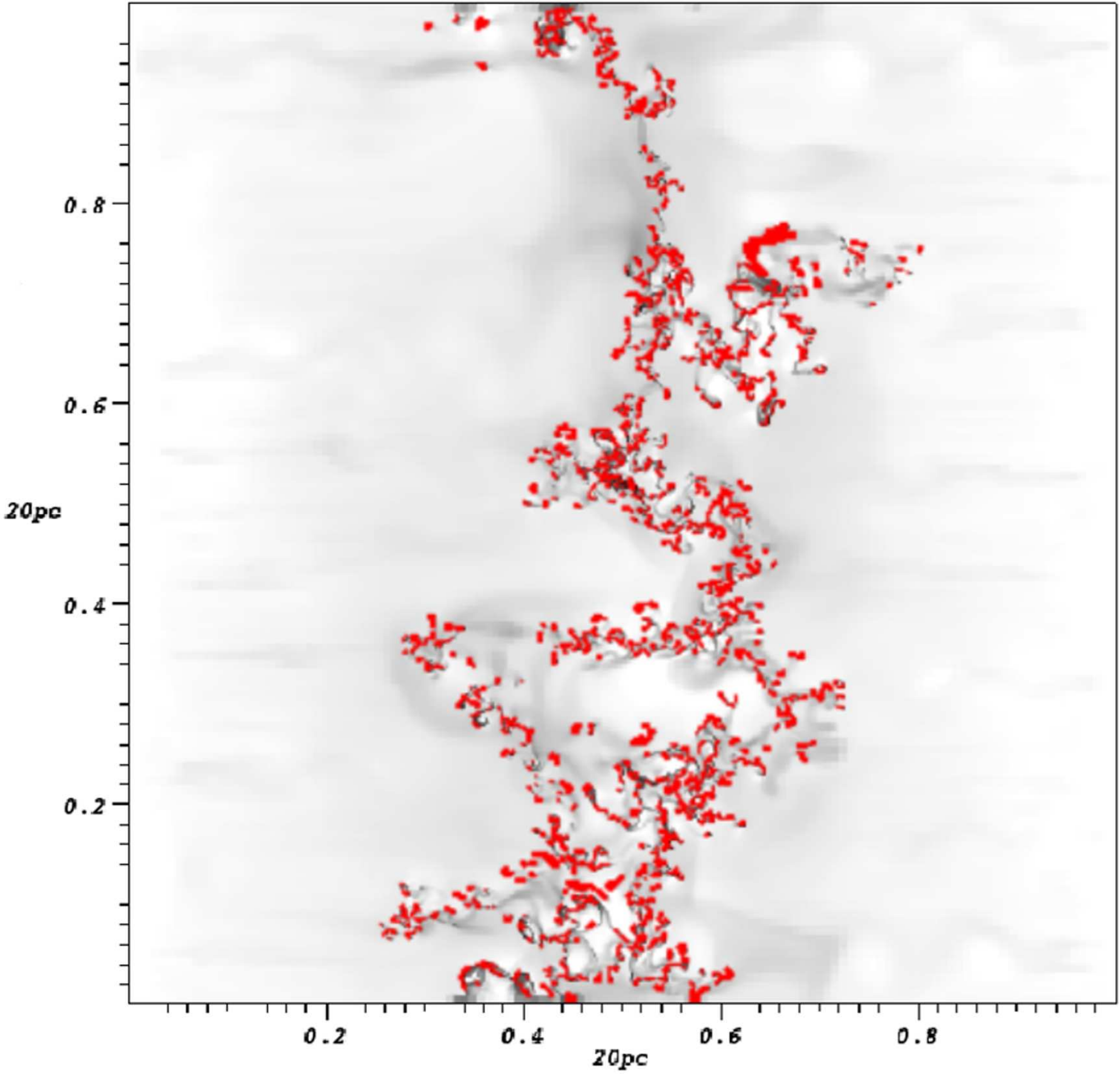}}
   \hspace{5mm}
   \subfloat[RCEN]{\label{fig:clumpsRCEN}\includegraphics[width=4cm]{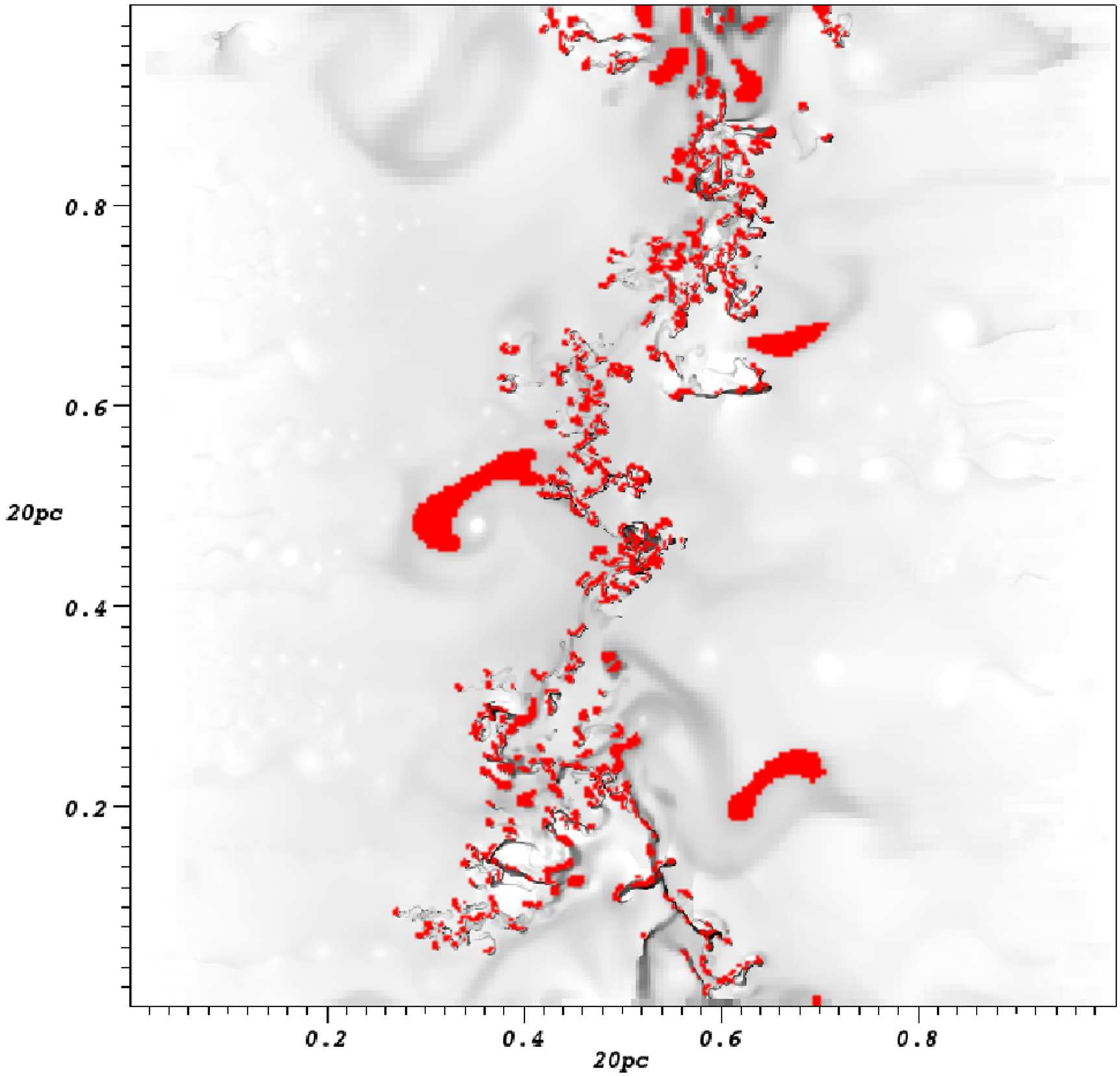}} \\
   \subfloat[Static]{\label{fig:fractalstatic}\includegraphics[width=4cm]{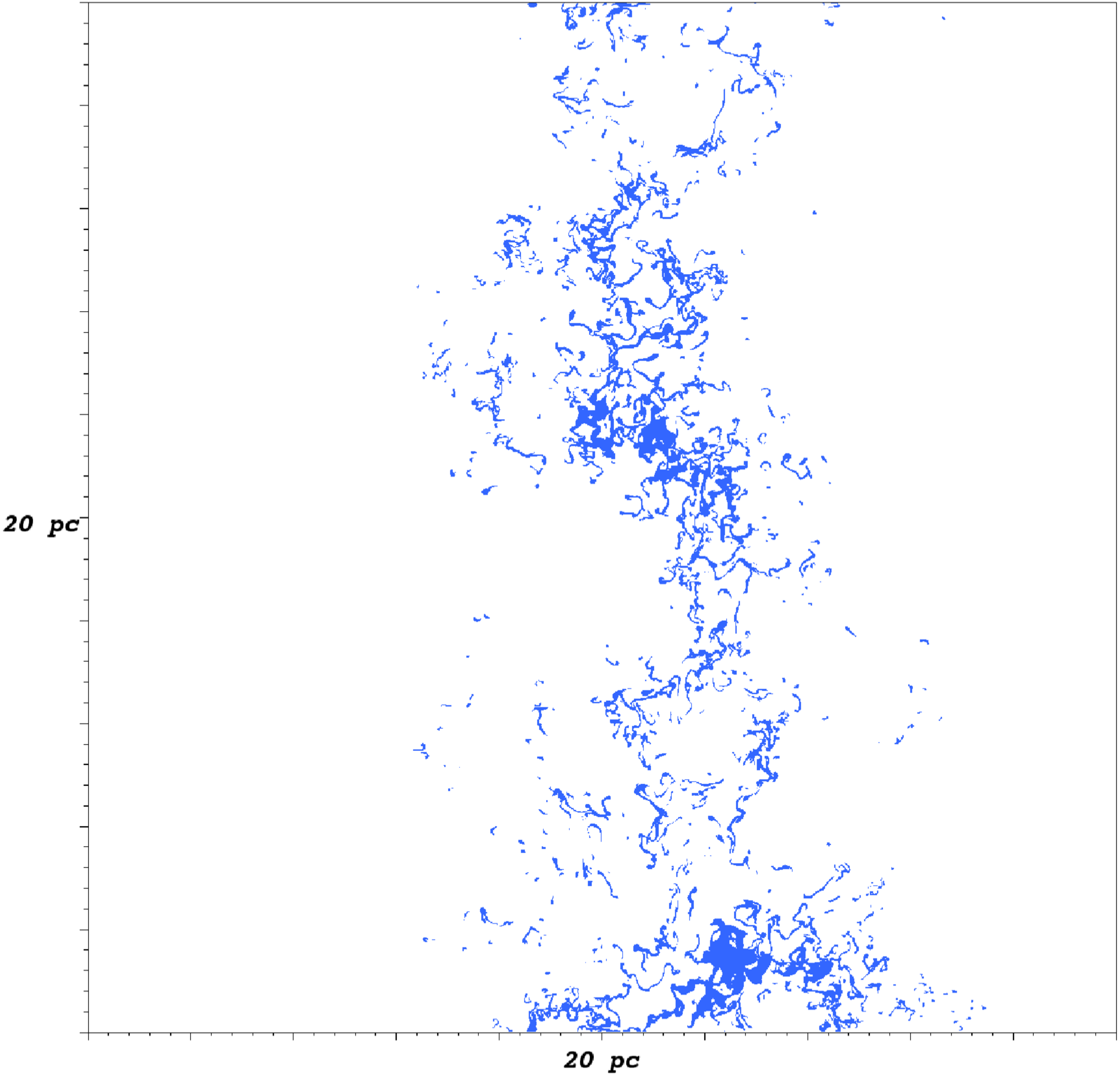}}
   \hspace{5mm}
   \subfloat[OD]{\label{fig:fractalOD}\includegraphics[width=4cm]{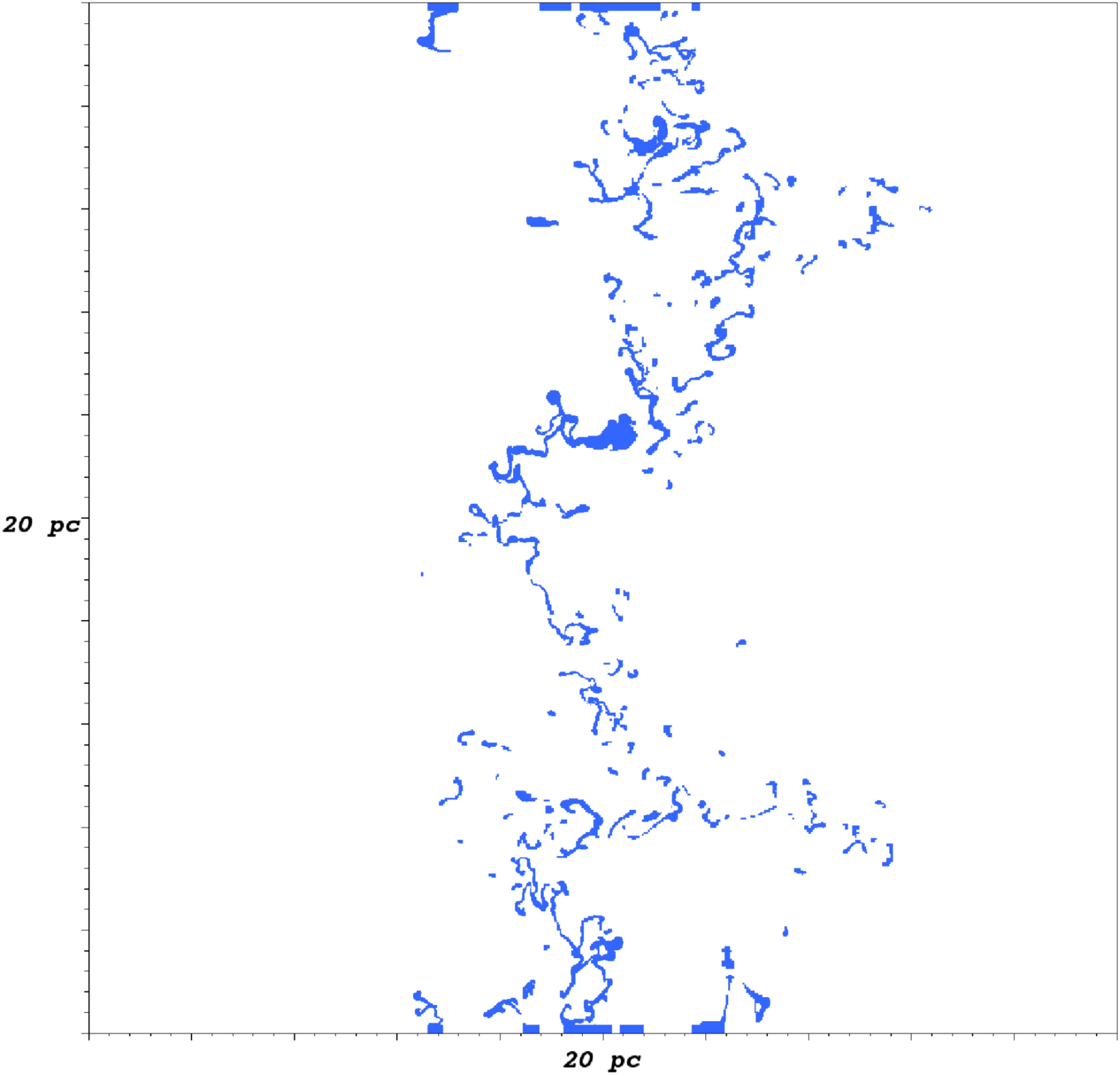}}
   \hspace{5mm}
   \subfloat[CT]{\label{fig:fractalCT}\includegraphics[width=4cm]{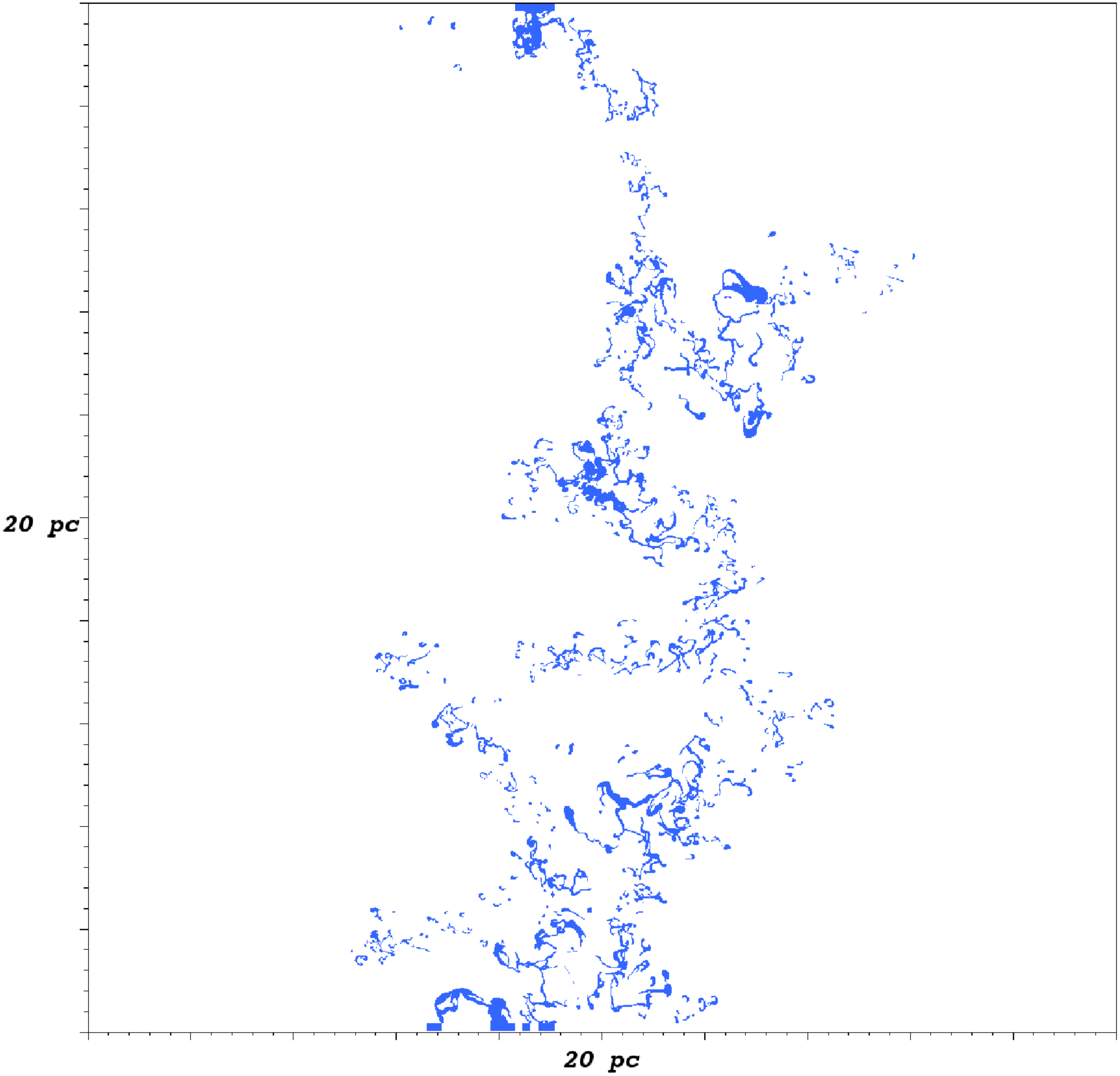}}
   \hspace{5mm}
   \subfloat[RCEN]{\label{fig:fractalRCEN}\includegraphics[width=4
cm]{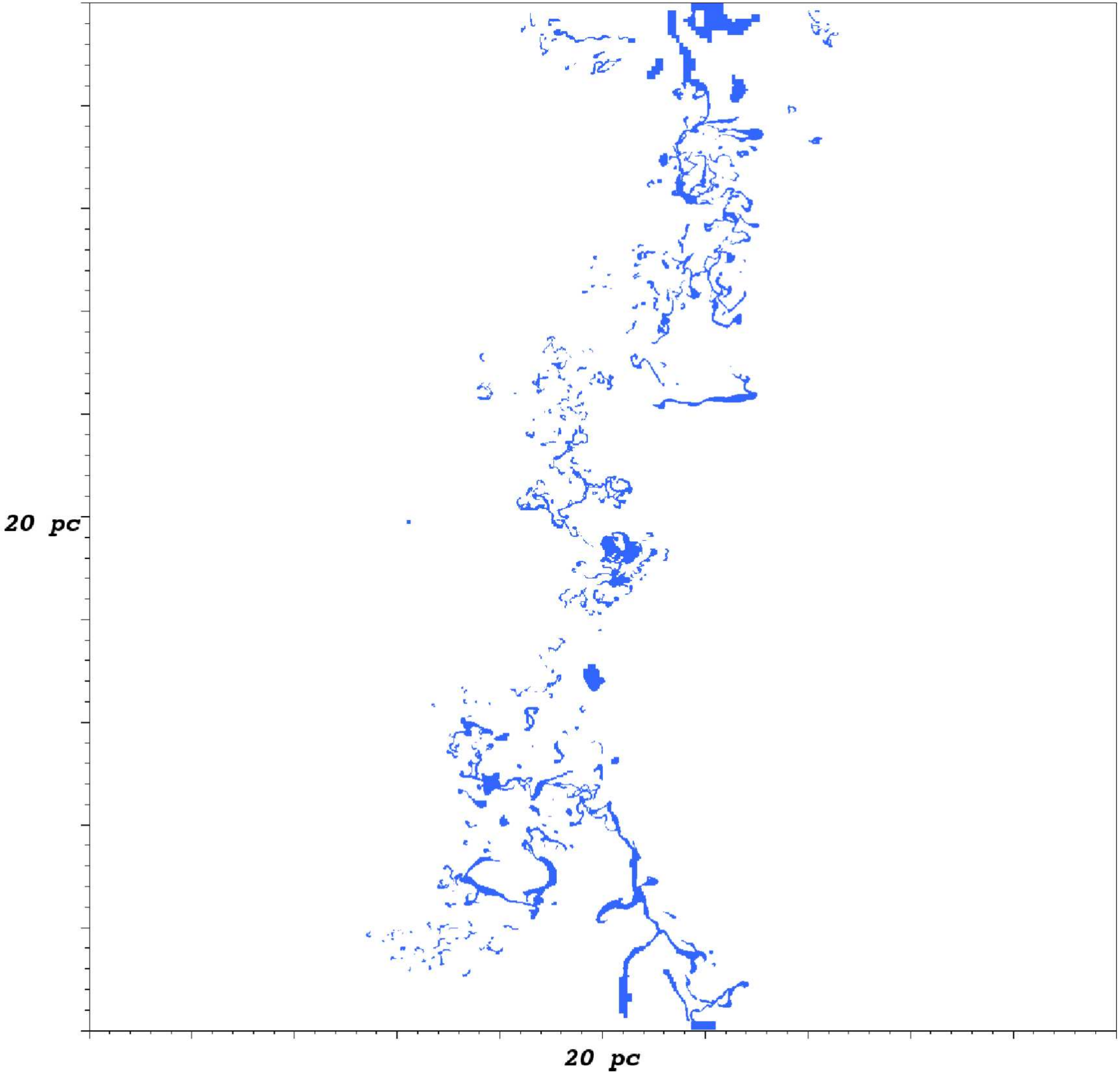}}
      \caption{Clump distributions (upper panel) and gas with number density $n=20\, \mathrm{cm}^{-3}$ (lower panel) for the AMR simulations compared to the static grid simulation. The clumps (displayed in red in the online version) were identified by a clumpfind algorithm and are superimposed on a grayscale density plot.}
         \label{fig:clumpfrac}
   \end{figure*}

The gravitationally unstable clumps of gas identified by the clumpfind algorithm in the static-grid
simulation at time $t=5\,\mathrm{Myrs}$ are depicted in Figure \ref{fig:clumpsstatic}. The corresponding results for the AMR runs are shown in Figures \ref{fig:clumpsOD}--\ref{fig:clumpsRCEN}. Table \ref{tab:clumps} lists the total number and the mean size of the clumps for each simulation. Also listed are the fractal dimensions of the gas regions with number density $n\ge 20$ cm$^{-3}$, which are plotted in Figures~\ref{fig:fractalstatic}-\ref{fig:fractalRCEN}. 

   \begin{table}[h]
\caption{Number $N$ and average size $\langle d\rangle$ of the clumps in the CNM; fractal dimension $D$ of gas with number density greater than $20$ cm$^{-3}$.}             
\label{tab:clumps}      
\bigskip
\centering                          
\begin{tabular}{l|rrr}        
\hline               
run & $N$ & $\langle d\rangle$ & $D$ \\    
\hline\hline                          
   static & $1025$ & $44$ & $1.45$ \\      
   OD &  $505$  & $102$ & $1.34$ \\
   OD-3 &  $481$  & $128$ & $1.33$ \\
   OD-4 &  $420$  & $157$ & $1.39$ \\
   CT &  $1239$  & $32$ & $1.50$ \\
   RCEN &  $1015$  & $85$ & $1.42$ \\
   \hline                                   
\end{tabular}
\end{table}
   
For refinement by OD, the fragmentation of the CNM is severely underestimated. The number of clumps is roughly half of the number in the static-grid simulation, and the clumps are typically larger. The lower degree of cold gas fragmentation results in a smaller fractal dimension (also see Figure \ref{fig:fractalOD}). If the criteria OD-3 and OD-4 are applied, the number of clumps decreases further, while their average size increases. In the case of criterion OD-4, a slightly higher fractal dimension is obtained, because the cold phase tends to fill broad, area-filling regions.
The cooling time criterion CT yields an amount of dense clumps with an average size that
compares well to the reference simulation (see Figure \ref{fig:clumpsCT}), although the degree of
fragmentation appears to be overestimated slightly. However, we found that this overestimation
decreases with the further evolution of the colliding flows and, thus, appears to be transient.
Refinement by RCEN also reproduces the number of clumps and the fractal dimension of dense
gas very well. However, there are some anomalously big clumps, which contribute to an average clump size that is systematically too large. In the plot showing gas at density $n\ge 20$ cm$^{-3}$ (see Figure~\ref{fig:fractalRCEN}), on the other hand, such anomalous structures are not visible. 
Although refinement by RCEN does not overproduce gas in the cold phase (as one can see from the
excellent agreement of the density and temperature pdfs in Figures~\ref{fig:densityRCEN} and~\ref{fig:tempRCEN}), there appears to be a bias toward bigger clumps with this refinement method.

\begin{figure*}[th]
   \centering
   \subfloat[static grid simulation]{\label{fig:contvortstatic}\includegraphics[width=0.49\linewidth]{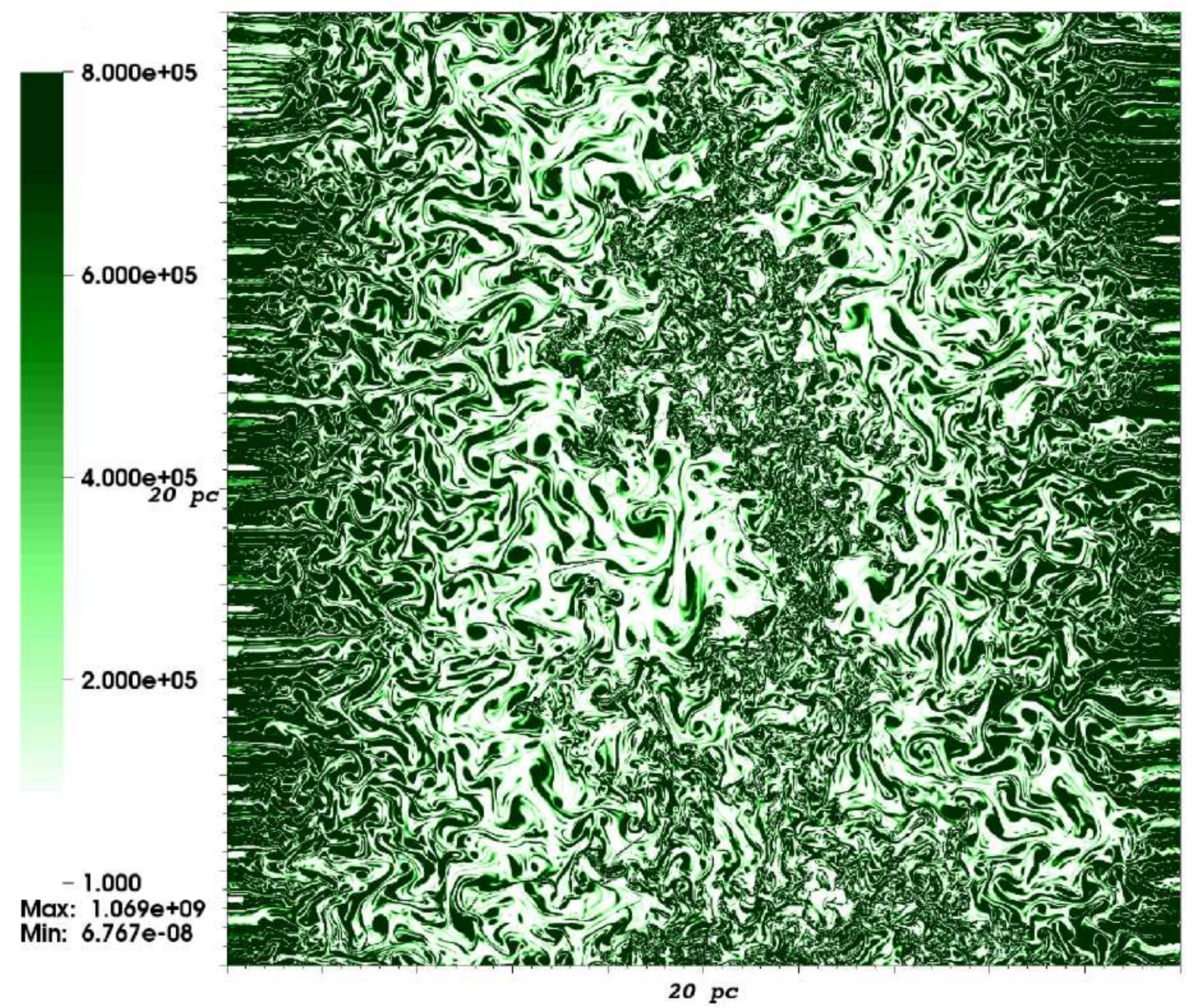}}
   \subfloat[AMR simulation (RCEN)]{\label{fig:contvortRCEN}\includegraphics[width=0.49\linewidth]{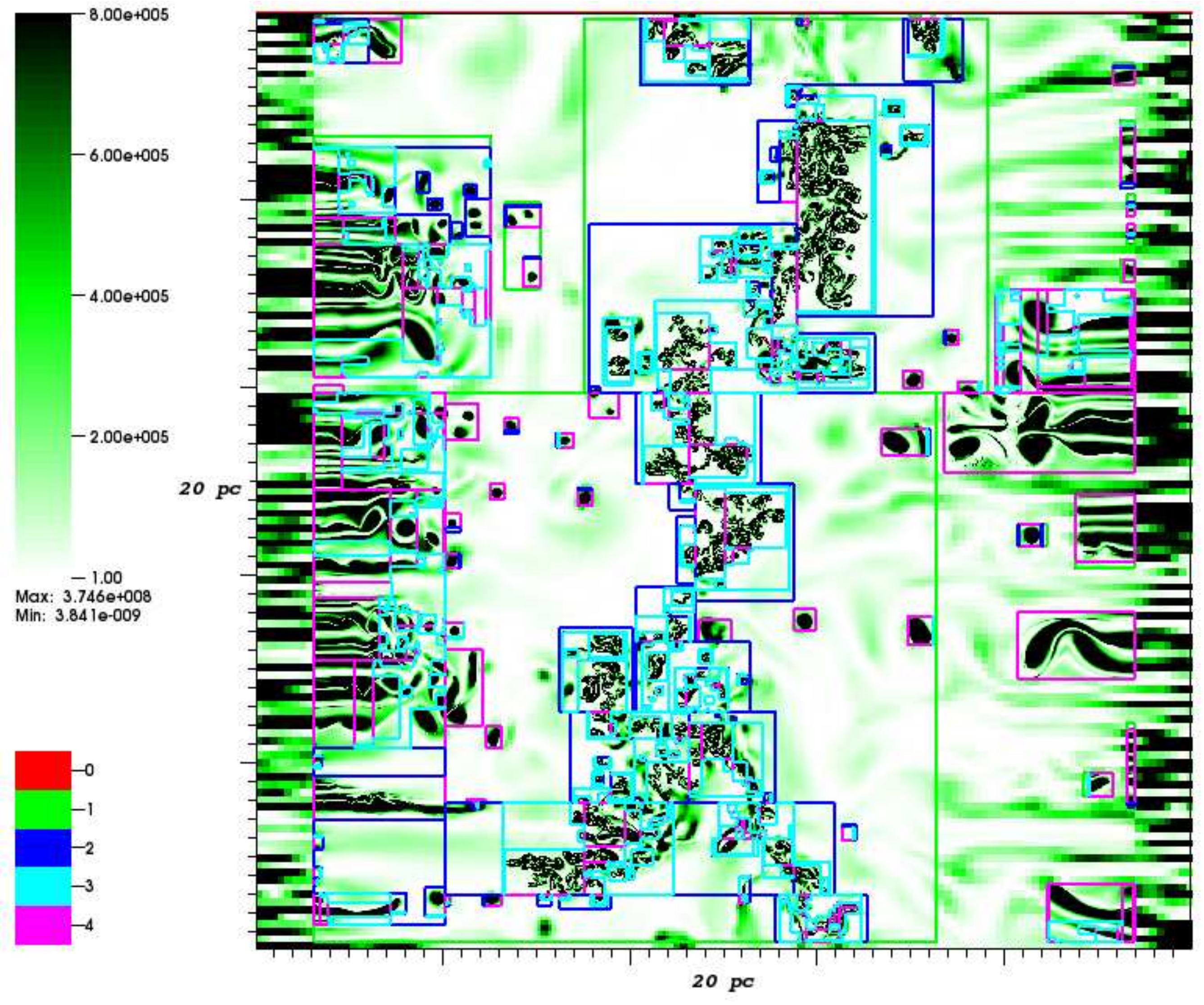}}
      \caption{Plots of the squared vorticity $\omega^2$ in the static-grid simulation (a) and
      	in the AMR simulation with refinement criterion RCEN (b), where the contours of the refinement levels are shown (color-coded in the online version).}
         \label{fig:contvortRCENstat}
   \end{figure*}

\begin{figure}[h]
   \centering
   \resizebox{\hsize}{!}{\includegraphics[width=\linewidth]{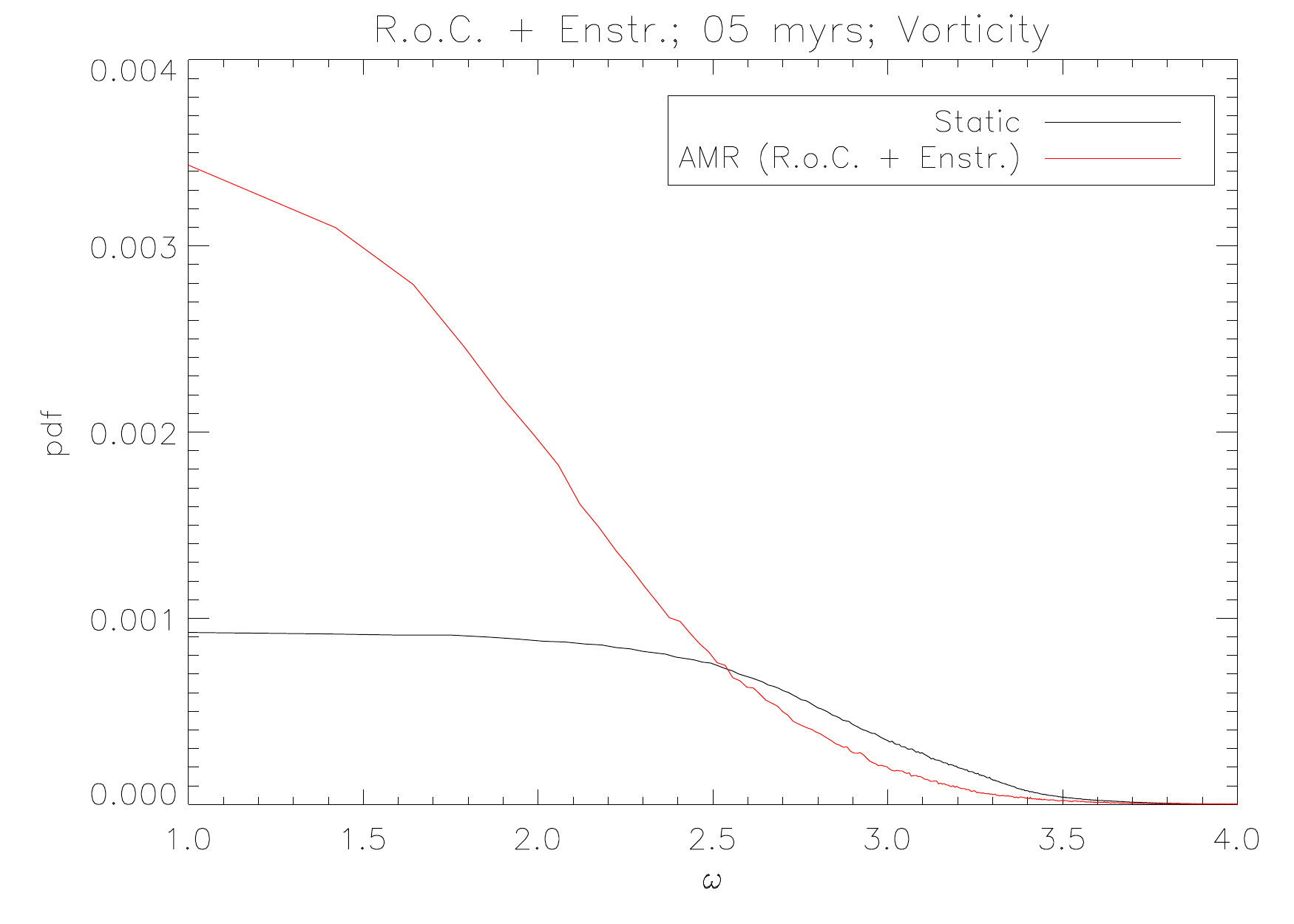}}
      \caption{Pdfs of the vorticity modulus $\omega$ for AMR-criterion RCEN (black curve) compared to the static grid simulation (red curve in the online version).}
         \label{fig:pdfvort}
   \end{figure}

In contrast to the phase separation and gas fragmentation, striking deviations of the turbulent
flow properties in the AMR vs. static grid simulations become apparent. Generally, a lot of turbulent small-scale structure is missing in the AMR simulations. Even for the criterion RCEN, which is based
on control variables related to turbulence, this is apparent from the contour plots of the squared
vorticity modulus shown in Figure~\ref{fig:contvortRCENstat}. Basically, the perturbations of the
velocity field imposed at the inflow boundaries are quickly smoothed out in AMR simulations, so that
turbulence is only produced by secondary (e.~g., Kelvin-Helmholtz) instabilities at the collision interface
in the central region of the computational domain. The reason is that all AMR cirteria, including RCEN,
select relatively large fluctuations, whereas smaller perturbations are suppressed. On a static grid, on the other hand, the perturbations are transported from the boundaries to the centre and actively contribute to the  production of turbulence. Consequently, small eddies are present in almost the whole domain in this case. Accordingly, the probability distribution of vorticity is markedly different (see Figure~\ref{fig:pdfvort}). In contrast, \citet{Sch09} found very close agreement of the vorticity pdfs in a static-grid and an AMR simulation with criterion RCEN for turbulence in a periodic box with large-scale forcing. Our results thus indicate that the merits of different refinement schemes  
are non-universal but rather depend on the properties of individual flow  
structures.

\section{Conclusions}

We performed two-dimensional simulations of colliding flows of warm atomic hydrogen with
a radiative cooling function as source term in the energy equation. The goal of our study
was the systematic comparison of AMR simulations, where different criteria for refinement
were applied, to a reference simulation on a static grid. 

While the probability distributions of mass density and temperature are well reproduced in
AMR simulations, regardless of the refinement technique, differences become apparent
in the fragmentation properties of the cold gas phase. As indicators, we used the total
number of clumps and their average size. The clumps were identified by a clumpfind
algorithm. In addition, we calculated the fractal dimension of dense gas, assuming a number density
threshold of $20$ cm$^{-3}$. Remarkably, the largest deviations from the clump statistics and
fractal dimension extracted from the static-grid simulation, were encountered for refinement by
overdensity, which is a commonly used refinement criterion in astrophysical AMR simulations. 
The deviations increase with the chosen density threshold. In this regard, it is important
to note that \citet{HenBan08} applied a density-based refinement criterion, where the
thresholds were chosen even higher than those considered in our study.
Good agreement, on the other hand, was obtained if the cooling time or enstrophy in combination
with the rate of compression (the negative rate of change of the velocity divergence) were
applied.

Substantial problems with AMR became apparent with regard to turbulent flow properties. Basically, none of our AMR runs were able to reproduce even remotely the small-scale structure of turbulence and the probability distributions of turbulent flow variables such as the vorticity modulus. This defficiency can be attributed to the selection effects introduced by adaptive techniques. The definition of thresholds for triggering refinement either selects strong local fluctuations (for example, large shear that gives rise to Kelvin-Helmholtz instabilities) or large-scale perturbations such as accumulations
of mass that become Jeans-unstable in self-gravitating gas. In this respect, the test problem
we investigated in this work is particularly tough, because turbulence stems from small-scale instabilities that are seeded by weak initial perturbations. The varying grid resolution in AMR simulations inevitably modulate the growth of these instabilities and, as a consequence, the production of turbulence is suppressed. This defficiency might be overcome by the application of a subgrid scale model, which transports turbulent energy contained in small eddies that are resolved on finer grids across coarser grid regions \citep[see][]{Maier09}.

The key point of using AMR is the computational cost for a given effective resolution. Indeed, Table \ref{tab:cputime} demonstrates that a substantial reduction of computation time is achieved with
AMR, especially if refinement by overdensity is applied. So, AMR is essentially a trade-off, where fast
computation has to be weighted carefully against the question whether the essential physics of the specific problem is captured. However, performing three-dimensional AMR simulations with very high effective resolution is conflicting. On the one hand the reduction of computation time will definitely be even greater, on the other hand there is the potential pitfall of inferring results that are properties of the numerics rather than the physics, since a comparison to a static-grid simulation is neither feasible nor desirable.

\begin{table}[h]
\caption{CPU time for the simulation runs.}             
\label{tab:cputime}      
\bigskip
\centering                          
\begin{tabular}{l|r}        
\hline    
   Method & CPU time \\    
\hline\hline                             
   static  &  $1172$ h $36$ min \\      
   OD &  $54$   h $20$ min \\
   CT & $225$ h $18$ min \\
   RCEN & $290$ h $57$ min \\
   \hline                                   
\end{tabular}
\end{table}

\begin{acknowledgements}
We thank Patrick Hennebelle and Edouard Audit for providing the cooling function that was used for this
numerical study. We also thank Paolo Padoan for sharing his clumpfind tool. Computations described in this work were performed using the Enzo code developed by the Laboratory for Computational Astrophysics at the University of California in San Diego (http://lca.ucsd.edu)
\end{acknowledgements}

\bibliographystyle{aa}
\bibliography{literature.bib}

\end{document}